\documentclass[%
 reprint,
superscriptaddress,
 amsmath,amssymb,
 aps,
 prx,
]{revtex4-2}

\usepackage{graphicx}
\usepackage{dcolumn}
\usepackage{bm}

\usepackage{bm}
\usepackage{amsfonts}
\usepackage{mathrsfs}
\usepackage[colorlinks, citecolor=red]{hyperref}

\begin{document}

\title{Multicell Atomic Quantum Memory as a\\Hardware-Efficient Quantum Repeater Node}
\author{C. Li$^{}$}
\altaffiliation[Present address: ISIS (UMR 7006), University of Strasbourg and CNRS, 67000 Strasbourg, France]{}
\email{c-l15@mails.tsinghua.edu.cn}
\affiliation{Center for Quantum Information, IIIS, Tsinghua University, Beijing 100084, PR China}

\author{S. Zhang$^{}$}
\affiliation{Center for Quantum Information, IIIS, Tsinghua University, Beijing 100084, PR China}

\author{Y.-K. Wu$^{}$}
\affiliation{Center for Quantum Information, IIIS, Tsinghua University, Beijing 100084, PR China}

\author{ N. Jiang$^{}$}
\affiliation{Department of Physics, Beijing Normal University, Beijing 100875, China}
\affiliation{Center for Quantum Information, IIIS, Tsinghua University, Beijing 100084, PR China}

\author{Y.-F. Pu$^{}$}
 \affiliation{Center for Quantum Information, IIIS, Tsinghua University, Beijing 100084, PR China}

\author{L.-M. Duan$^{\footnotemark[0]}$}
\email{lmduan@tsinghua.edu.cn}
\affiliation{Center for Quantum Information, IIIS, Tsinghua University, Beijing 100084, PR China}

\begin{abstract}
For scalable quantum communication and networks, a key step is to realize a quantum repeater node that can efficiently connect different segments of atom-photon entanglement using quantum memories. We report a compact and hardware-efficient realization of a quantum repeater node using a single atomic ensemble for multicell quantum memories. Millisecond lifetime is achieved for individual memory cells after suppressing the magnetic-field-induced inhomogeneous broadening and the atomic-motion-induced spin-wave dephasing. Based on these long-lived multicell memory cells, we achieve heralded asynchronous entanglement generation in two quantum repeater segments one after another and then an on-demand entanglement connection of these two repeater segments. As another application of the multicell atomic quantum memory, we further demonstrate storage and on-demand retrieval of heralded atomic spin-wave qubits by implementing a random access quantum memory with individual addressing capacity. This work provides a promising constituent for efficient realization of quantum repeaters for large-scale quantum networks.
\end{abstract}

\maketitle

\section{Introduction}

The realization of long-distance quantum communication and large-scale quantum networks is one of the primary goals in quantum information science \cite{kimble2008quantum,wehner2018quantum}.
Because photons, the ideal information carrier for quantum communication, are still subject to the intrinsic exponential loss in the optical fiber, the quantum repeater protocol is proposed to provide a more efficient scaling with the transmission distance \cite{briegel1998quantum}. In this protocol, the full quantum communication channel is divided into shorter segments, with entanglement first generated in the elementary segments and then extended to longer distances through entanglement swapping and purification in a divide-and-conquer manner. With the help of long-lived quantum memories, the quantum repeater protocol can achieve a polynomial overhead versus the distance compared with the exponential cost for direct transmission.

The Duan-Lukin-Cirac-Zoller (DLCZ) protocol is a well-known scheme to realize quantum repeaters \cite{duan2001long,RevModPhys.83.33}. It uses atomic ensembles as a convenient interface between flying photons for information transmission and atomic spin-wave excitations for quantum memory. Significant progress has been made for the ensemble-based quantum repeaters: fundamental building blocks such as the entanglement generation, storage, and retrieval have been achieved  \cite{PhysRevLett.95.040405,choi2008mapping}; entanglement distribution between distant ensembles has been realized \cite{chou2005measurement, Chou1316,yuan2008experimental,yu2020entanglement}; and recently the scaling-changing behavior of the entanglement connection has also been demonstrated \cite{pu2021experimental}.

In the DLCZ protocol, two atomic ensembles are used in a quantum repeater node to hold entanglement with the two adjacent links to be connected. With the development of multicell atomic quantum memories, it becomes possible to realize a quantum repeater node using several long-lived memory cells inside a single atomic ensemble. The multicell quantum memory has first been proposed as a variant of the DLCZ protocol to improve the entanglement distribution rate \cite{collins2007multiplexed, PhysRevLett.98.190503}: by generating entanglement in multiple pairs of memory cells and connecting them in a multiplexed way, an improved communication rate and a less stringent requirement on the memory lifetime can be achieved. Previously, multicell quantum memories using the spatial \cite{lan2009multiplexed, parniak2017wavevector,  Nicolas2014, zhang2016experimental, wen2019multiplexed, Langenfeld2020experimental}, temporal \cite{usmani2010mapping, kutluer2017solid, tiranov2017quantification, PhysRevApplied.12.024062} and spectral modes \cite{PhysRevLett.113.053603, PhysRevLett.123.080502} are realized in atomic ensembles \cite{wen2019multiplexed, zhang2016experimental, parniak2017wavevector, Nicolas2014, lan2009multiplexed}, solid state spins \cite{usmani2010mapping, tiranov2017quantification, kutluer2017solid} and cavity QED systems \cite{PhysRevLett.123.080502, PhysRevApplied.12.024062, Langenfeld2020experimental, PhysRevLett.113.053603}. However, a quantum repeater node in a single multicell quantum memory has not been realized yet due to the experimental difficulty in combining multimode storage capacity, individual addressing, and long coherence time in a single system.

\begin{figure*}[!tbp]
  \centering
  \includegraphics[width=16cm]{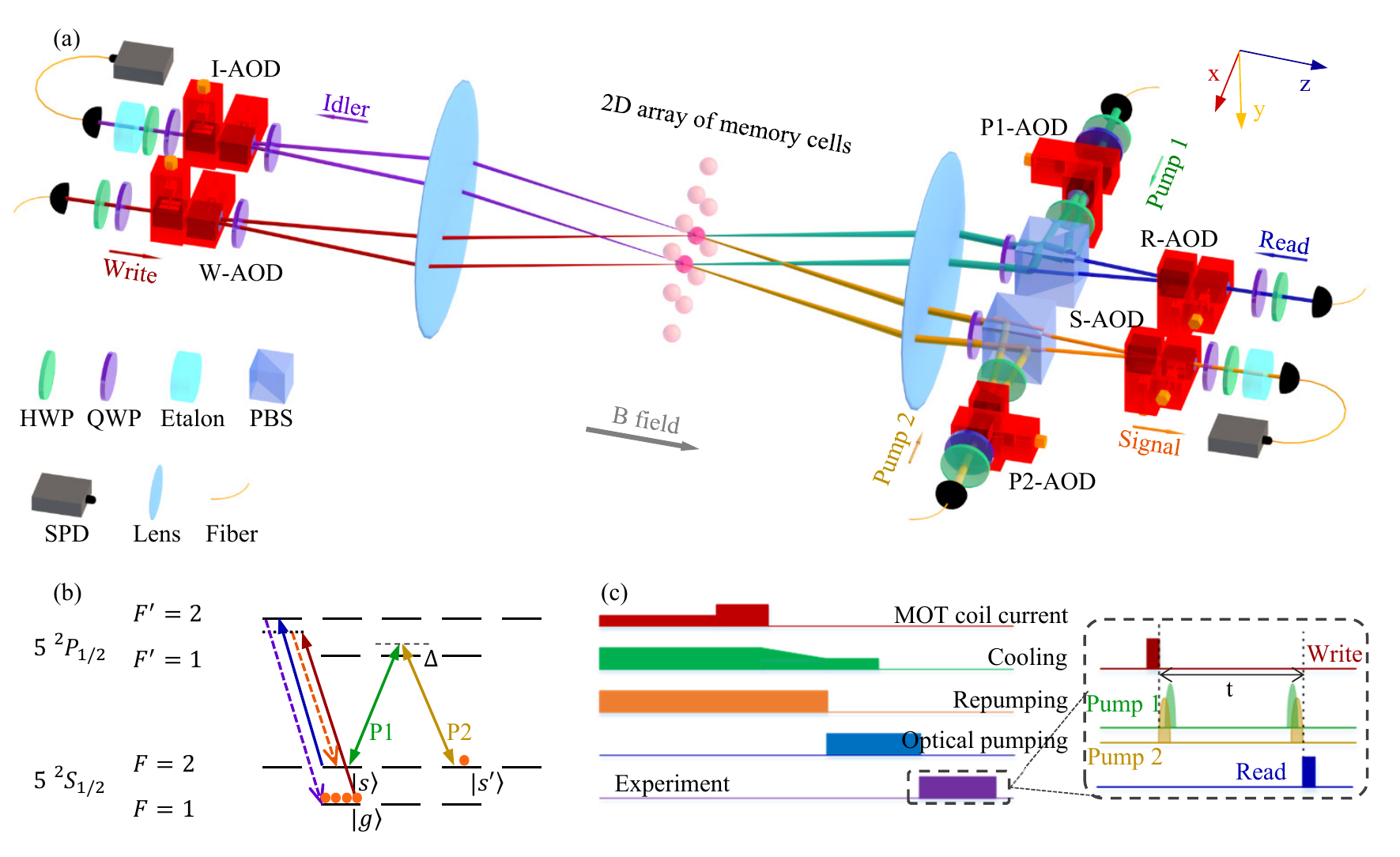}\\
  \caption{\textbf{Experiment setup for the long-lived MAQM.} \textbf{(a)} Simplified setup. HWP represents half-wave plates, QWP for quarter-wave plates, PBS for polarization beam splitters and SPD for single photon detectors. The angles between write/idler beams and read/signal beams are the same $1.5^{\circ}$. The Gaussian sizes of the signal/idler, write/read, Pump1/Pump 2 modes are $65\, \mu$m, $125\,\mu$m and $275\,\mu$m respectively. For clarity, $3\times 4$ memory cells are sketched in the figure, which are the selected regions from the cold ensemble by the orthogonally placed AODs. The distance between two adjacent memory cells is $420\,\mu$m. The Fabry-Perot cavities (etalons) are used to filter the leakage pulses from the signal and idler modes. The bias magnetic field $B$ is used to define the quantization axis. Two cells are addressed to manifest the entanglement generation between the signal photon and the spin-wave excitation in two spatial modes. More details about the setup can be found in Appendix~A. \textbf{(b)} Energy diagram. The ensemble is initially prepared on the ground state $|g\rangle$. The storage states, $|s\rangle$ and $|s^\prime\rangle$, are used to store the spin-wave excitation. A write pulse (red solid arrow) is applied to generate a signal photon (orange dashed arrow) and a spin wave. The spin wave is transferred by two Raman pumping beams, $P1$ (green solid arrow) and $P2$ (gold solid arrow), and it is finally retrieved by a read pulse (blue solid arrow) into the idler photon (purple dashed arrow). The write beam is red detuned by $28\,$MHz, and the detuning of $P1$ and $P2$ is $40\,$MHz. \textbf{(c)} Experimental sequence. The ensemble in MOT is compressed, polarization gradient cooled and further optical pumped to the ground state before the experiment. The circled part is one complete write-in and read-out cycle with a storage time $t$ if a signal photon is detected. The Pump 1 and Pump 2 pulses are both in a temporal Gaussian profile (see Appendix~B for details).}
\end{figure*}

Here, we experimentally demonstrate a quantum repeater node in a multicell atomic quantum memory (MAQM). The lifetime of each memory cell is extended to the scale of one millisecond after suppressing the inhomogeneous broadening induced by the magnetic field gradient and the atomic-motion-induced dephasing. With the achieved long lifetime, two segments of atom-photon entanglement can be generated asynchronously and stored into different memory cells of the same MAQM. We then demonstrate entanglement connection of these two repeater segments on the quantum repeater node realized with a single MAQM by projecting the stored spin wave modes onto a Bell state. Compared with the realization of the quantum repeater node with two distinct atomic ensembles \cite{pu2021experimental}, this realization is more compact and hardware-efficient, less vulnerable to the environmental noise, and more convenient for manipulation of information stored in different memory cells. As another application of the MAQM with long coherence time, we further demonstrate a random access quantum memory (RAQM) for states of heralded spin wave excitations through individual addressing of different memory cells. Compared with the previous implementation of the RAQM \cite{jiang2019experimental,Langenfeld2020experimental}, here we can make use of the achieved long memory time to herald excitations stored into the MAQM by successfully detecting a corresponding photon. Our experiment provides a compact and hardware-efficient implementation of the quantum repeater node, which is a key element for future large-scale quantum repeaters and quantum networks.

\section{Experimental Results}
\subsection{Improving the memory cell lifetime}

In the DLCZ protocol, a signal photon detection heralds a collective spin-wave excitation stored in the atomic ensemble \cite{duan2001long}, which is described as
\begin{equation}
\label{eq1}
|\Psi\rangle_{gs}=\frac{1}{\sqrt{N}}\sum_{j=1}^N e^{i\boldsymbol{k_s\cdot r_j}}|g_1 \cdots s_j \cdots g_N\rangle ,
\end{equation}
where $N$ is the total atom number of the ensemble, $|g\rangle$ and $|s\rangle$ are the ground state and the storage state for the atoms respectively, the subscript $gs$ represents a single excited spin wave from the $|g\rangle$ to the $|s\rangle$ state, $\boldsymbol{r_j}$ is the position of the $j$-th atom, and $\boldsymbol{k_s}$ is the  momentum of the spin wave. There are two major limitations for the lifetime of the spin wave. One is the inhomogeneous broadening of the spin transition caused by the magnetic field gradient. A solution is to pick the ``clock states'' as the initial and the storage states \cite{PhysRevA.66.053616}, which is first-order insensitive to the magnetic field. The other one is the random atomic motion. It induces random phases for different atoms during the storage period and destructs the coherence of the spin wave. Several methods can suppress the atomic-motion-induced dephasing: the collinear configuration \cite{Bao2012} or near collinear configuration \cite{Long2020wang} can be used to reduce the momentum $\boldsymbol{k_s}$ of the spin wave; loading atoms into an optical dipole trap can confine the atomic motion alongside the momentum direction \cite{LongZhao2009}; applying spin echoes can eliminate the accumulated random phases during the storage \cite{PhysRevLett.115.133002}.

In this experiment, we use the spin wave freezing method \cite{PhysRevA.93.063819} to eliminate the momentum of the spin wave, thus freezing the motion-induced decoherence. The experimental setup is depicted in Fig.~1a. The addressing system with pairs of orthogonally placed acousto-optic deflectors (AODs) is used to create an MAQM in a cold $^{87}$Rb ensemble (see details in Appendix~A and Refs.~\cite{pu2017experimental,jiang2019experimental}). As shown in Fig.~1b, the ensemble is initially prepared to the ground state $|g\rangle=|F=1,\, m_F=-1\rangle$. A weak write pulse probabilistically creates a signal photon and a correlated spin-wave excitation $|\Psi\rangle_{gs}$ [Eq.~(\ref{eq1})] via spontaneous Raman scattering with $|s\rangle=|F=2,\, m_F=-1\rangle$. The momentum of the spin wave is determined by the directions of the write pulse $\boldsymbol{k}_W$ and the detected signal photon $\boldsymbol{k}_S$ as $\boldsymbol{k_s}=\boldsymbol{k}_W-\boldsymbol{k}_S$. After that, two Raman pumping beams (Pump 1 and Pump 2) are applied to transfer the spin wave to the state  $|\Psi\rangle_{gs^\prime}$, where $|s^\prime\rangle=|F=2,\, m_F=+1\rangle$. In this way, the momentum of the spin wave is converted to $\boldsymbol{k_{s^\prime}}=\boldsymbol{k_s}+\boldsymbol{k}_{P_1}-\boldsymbol{k}_{P_2} $. An elaborate arrangement of the Raman pumping beams ensures that the final spin-wave momentum is wiped out, hence the motion-induced dephasing is suppressed; meanwhile, the usage of the clock states, $|g\rangle$ and $|s^\prime\rangle$, suppresses the inhomogeneous broadening induced by the magnetic field \cite{PhysRevA.81.041805}. After a storage time $t$,  the spin wave is pumped back to the state $|\Psi\rangle_{gs}$, and the momentum $\boldsymbol{k_s}$ is recovered. Then the spin wave can be retrieved in the phase-matching direction by a strong read pulse. To transfer the spin wave between the two states, $|\Psi\rangle_{gs}$ and $|\Psi\rangle_{gs^\prime}$, the stimulated rapid adiabatic passage (STIRAP) technique is applied, which is robust to the fluctuation of the laser intensity and the variation of the beam overlapping \cite{Bergmann_2019}.

\begin{figure}[htb]
  \centering
  \includegraphics[width=8.6cm]{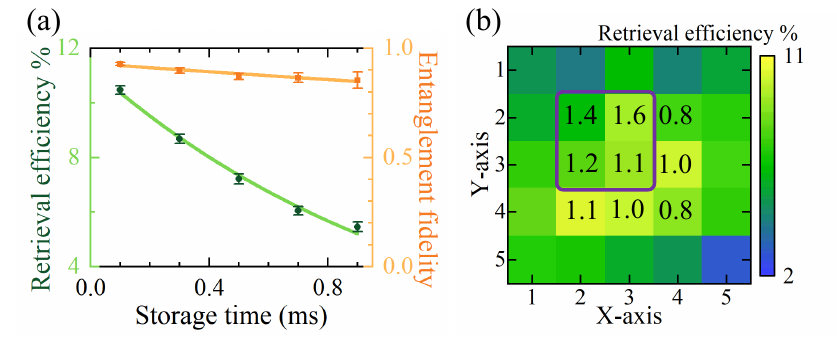}\\
  \caption{\textbf{MAQM with long lifetime.} \textbf{(a)} The retrivel efficiency of the central cell (green dot), and the entanglement fidelity for a path qubit encoded in the two central cells with the corresponding signal photon (orange square) versus the storage time $t$. The error bars are calculated by the Monte-Carlo simulation with a Poisson distribution assumption for photon counts. The lifetime of the cell is fitted from an exponential function. \textbf{(b)} The color map of the retrieval efficiency after $t=0.1\,$ms and the fitted storage lifetime (the number in each cell in the unit of millisecond). The circled four cells are used to generate multiple spin waves in the following experiments.}
\end{figure}

The retrieval efficiency and the lifetime of the memory cells are shown in Fig.~2. The $1/e$ lifetimes of the nine central cells are fitted from the exponential decay of the retrieval efficiency versus the storage time, and all the lifetimes are on the order of one millisecond. We also verify the preservation of entanglement during the spin wave transfer and storage processes. Two central cells, $(x_2, y_3)$ and $(x_3, y_3)$, are chosen to generate the entangled state between a path-encoded signal photon and a spin wave excitation as (with the vacuum and higher order terms neglected)
\begin{equation}
\label{eq2}
\frac{1}{\sqrt{2}}(|x_2,y_3\rangle_{s}|x_2,y_3\rangle_{a}+|x_3,y_3\rangle_{s}|x_3,y_3\rangle_{a}),
\end{equation}
where $|x_i,y_3\rangle_{s}$ and $|x_i,y_3\rangle_{a}$ $(i =2, 3)$ denote the signal photon and the spin wave generated in the cell $(x_i, y_3)$, respectively. The entanglement fidelity is calculated from the density matrix reconstructed by quantum state tomography \cite{PhysRevA.64.052312}. As shown in Fig.~2a, the fidelities are above $80\%$ for the storage time up to $0.9\,$ms, which proves that the coherence between two memory cells, and the quantum correlation between the signal photon and the spin wave, are both well maintained.
Here we use two neighboring cells to encode the qubit so as to provide an upper bound on the crosstalk error due to the finite laser beam width, although as we discuss in Appendix~C, in this experiment the dominant crosstalk error may come from the imperfect STIRAP pulses.

\subsection{Quantum repeater node in a multicell atomic quantum memory}

\begin{figure}[!tbp]
  \centering
  \includegraphics[width=8.6cm]{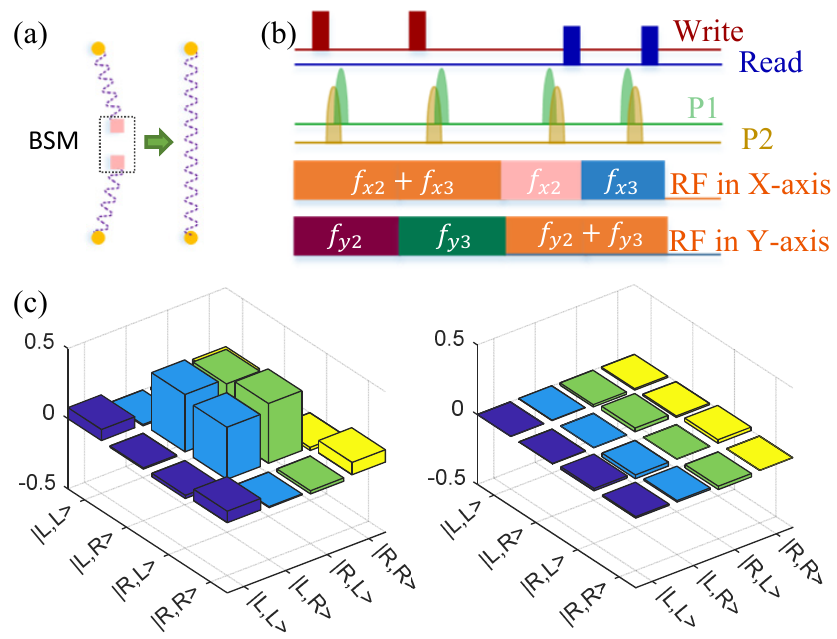}\\
  \caption{\textbf{Entanglement connection at a quantum repeater node.} \textbf{(a)} The scheme of the entanglement connection. A successful Bell state measurement (BSM) of two spin waves heralds the connection of two adjacent quantum repeater links. \textbf{(b)} Experimental sequence. After two entangled states are generated in cell pairs 1 and 2, $10\,\mu$s is waited to project the $L$ modes of the spin waves onto the state $|\phi_L\rangle$, and another $5\,\mu$s to project the $R$ modes onto the state $|\phi_R\rangle$. \textbf{(c)} The reconstructed density matrix of the two signal photons. $L$ and $R$ represent the signal photons detected from the corresponding memory cells of the two pairs. The signal photon detection counts are recorded only when the two idler photons are also detected. The measured state fidelity is $76.6\pm3.9\%$.
  }
\end{figure}

To demonstrate the functioning of a quantum repeater node, we follow the idea of Ref.~\cite{pu2021experimental} to consider a primitive quantum repeater with two segments. The atom-photon entanglements in the two segments are created one after another heralded by the detection of the signal photons (note that since the two photons do not need further connection here, they can be measured prior to the entanglement swapping step \cite{pu2021experimental}). Then we retrieve the two spin-wave qubits and project them onto a Bell state to perform an entanglement swapping.

Specifically, we choose two pairs of memory cells to generate and to store the two spin waves (cells $(x_2,\,y_2)$ and $(x_3,\,y_2)$ as pair 1 and cells $(x_2,\,y_3)$ and $(x_3,\,y_3)$ as pair 2) as shown in Fig.~2b. For simplicity, here we label the two spatial modes as $|L\rangle$ and $|R\rangle$ in each pair. The storage lifetime on the order of milliseconds allows us to create two spin wave excitations one by one, which are typically separated by hundreds of microseconds due to the relatively low excitation probability.
In this way, we generate two pairs of atom-photon entangled states as
\begin{equation}
\begin{aligned}
|\Psi\rangle_{s_1 a_1 s_2 a_2} \equiv\frac{1}{2}&(|L\rangle_{s_1}|L\rangle_{a_1}+|R\rangle_{s_1}|R\rangle_{a_1})\\
\otimes&(|L\rangle_{s_2}|L\rangle_{a_2}+|R\rangle_{s_2}|R\rangle_{a_2}).
\end{aligned}
\end{equation}

The Bell state measurement between the two spin waves is achieved using a similar scheme as in the DLCZ protocol \cite{duan2001long,RevModPhys.83.33}. First we retrieve the spin waves in the $L$ modes of the two pairs with equal amplitude, which can be understood as a projection onto the state $|\phi_L\rangle=(|L\rangle_{a_1}+|L\rangle_{a_2})/\sqrt{2}$. Then we similarly retrieve the spin waves in the $R$ modes for a projection onto $|\phi_R\rangle=(|R\rangle_{a_1}+|R\rangle_{a_2})/\sqrt{2}$. Conditioned on two coincident detection events of idler photons from these two cases, we achieve a successful projection on the Bell state $(|L\rangle_{a_1}|R\rangle_{a_2}+|R\rangle_{a_1}|L\rangle_{a_2})/\sqrt{2}$ given that our initial state $|\Psi\rangle_{s_1 a_1 s_2 a_2} $ has only one excitation in $a_1$ and one excitation in $a_2$. The corresponding signal photons are hence projected to the entangled state $(|L\rangle_{s_1}|R\rangle_{s_2}+|R\rangle_{s_1}|L\rangle_{s_2})/\sqrt{2}$. In Fig.~3c we reconstruct the density matrix of this signal-signal entangled state by quantum state tomography. The measured state fidelity is $(76.6\pm3.9)\%$, which is above the classical threshold of 1/2 and thus verifies the successful entanglement swapping. A detailed analysis about the limiting factors for the entanglement fidelity are shown in Appendix~D.

\subsection{Random access quantum memory}

\begin{figure}[!tbp]
  \centering
  \includegraphics[width=8.6cm]{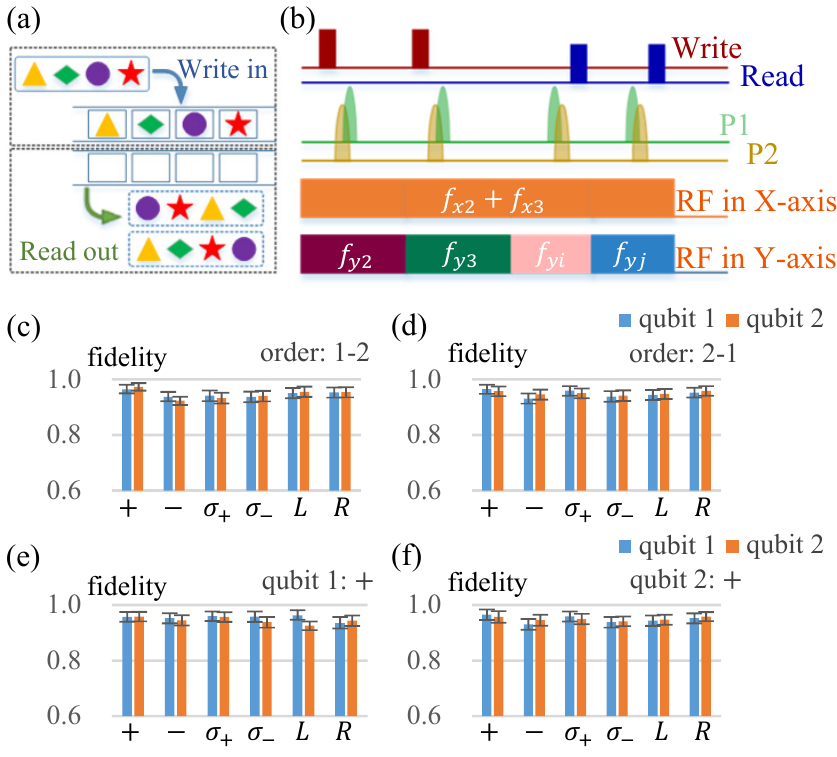}\\
  \caption{\textbf{Random access quantum memory.} \textbf{(a)} An illustration of an RAQM. Multiple memory cells are used to store the qubits, and the qubits can be read out in an arbitrary order. \textbf{(b)} Time sequence for the experiment. First the cell pair 1 is addressed to generate the spin wave. Once a signal photon is detected, a $10\,\mu$s time interval is waited to switch all the AODs to address cell pair 2. Then after the spin wave is generated in the second cell pair, $10\,\mu$s is waited to read out the qubit $i$ $(i=1,2)$ and another $5\,\mu$s to read out the other qubit $j$ $(j \ne i)$. We consider six complementary input states $|L\rangle$, $|R\rangle$, $|\pm\rangle=(|L\rangle\pm|R\rangle)/\sqrt{2}$, $|\sigma_{\pm}\rangle=(|L\rangle\pm i|R\rangle)/\sqrt{2}$ for each qubit. In \textbf{(c, d)}, we prepare the two qubits in the same states and read them out in the order of qubits 1-2 and qubits 2-1 respectively to measure the fidelity. The qubits in different states are also stored as shown in \textbf{(e)} (fix qubit 1 to $|+\rangle$ and vary qubit 2) and in \textbf{(f)} (fix qubit 2 to $|+\rangle$ and vary qubit 1). The state fidelity is calculated by reconstructing the density matrix through quantum state tomography. The error bars denote one standard deviation.}
\end{figure}

As another application of the multimode storage and individual addressing capability, we further demonstrate an RAQM in our setup. An RAQM can store a series of qubit states and then read them out on demand in an arbitrary order (see Fig.~4a). Unlike previous works \cite{jiang2019experimental, Langenfeld2020experimental}, here we are able to create multiple spin waves heralded by signal photon detections and then retrieve them on demand. Consider the same two pairs of memory cells as before with qubit information encoded in the $|L\rangle$/$|R\rangle$ paths. By setting the relative amplitudes and phases of the write beams on the two paths using the W-AODs, we get an entangled state $\cos{\theta}|L\rangle_{s}|L\rangle_{a}+e^{i\phi}\sin{\theta}|R\rangle_{s}|R\rangle_{a}$ similar to Eq.~(\ref{eq2}).  If the detection basis of the signal photon is set to $(|L\rangle_s+|R\rangle_s)/\sqrt{2}$, the spin wave stored in the cells is projected to $ \cos{\theta}|L\rangle_{a}+e^{i\phi}\sin{\theta}|R\rangle_{a}$.
Hence, the spin-wave qubit is controlled by the W-AODs, and its creation is heralded by the signal photon detection. Once two spin waves are generated one by one, they can be read out in a programmable way after a controllable storage time (sequences shown in Fig.~4b) in an arbitrary order. To characterize the performance of the RAQM, we store two qubits in various initial states and measure the state fidelity of the retrieved qubits through quantum state tomography (Fig.~4c-f). In all these cases, the fidelity exceeds the classical bound of $F_B=2/3$ \cite{jiang2019experimental} and thus shows the quantum nature of the memory.

\section{Discussion and Conclusion}

Currently in our experiment, the memory lifetime is limited by the atomic free expansion, which can be mitigated by loading the micro-ensembles into a dipole trap array \cite{Wang2020Preparation}. Our setup can also be integrated with an optical cavity to increase the retrieval efficiency \cite{Bao2012, PhysRevLett.123.263601}.  Toward a practical quantum repeater node, one can use telecom photons \cite{PhysRevX.9.041033,radnaev2010quantum} for lower transmission loss at long distance. Moreover, our MAQM can hold more than two qubit modes, thus supports more complicated quantum network patterns than the linear configuration in the original quantum repeater protocol.

Although in this work we focus on demonstrating a compact quantum repeater node within a single atomic ensemble, our setup with long memory lifetime, multimode storage and individual addressing may also find applications in a multiplexed quantum repeater \cite{collins2007multiplexed}. However, currently only one spatial mode can be detected at one time with our multiplexing and demultiplexing optical circuits using AODs, which prevents the demonstration of an enhanced entanglement generation rate. Therefore, to realize a multiplexed quantum repeater, new facilities are required to detect multiple spatial modes simultaneously, for example using CMOS single photon detector \cite{parniak2017wavevector} or photomultipliers array coupled with a multi-core fiber or a fiber array \cite{debnath2016demonstration,Ding2017high,hu2020efficient}.

Our work may also find applications in neutral-atom-based quantum computers \cite{Saffman_2016}, where Rydberg interaction generates entanglement between nearby atomic qubits \cite{PhysRevLett.115.093601}. Then the atom-photon interface we demonstrate in this work and the Bell state measurement can be used to entangle the atomic qubits in distant memory cells or even in different ensembles, thus significantly increases the connectivity of the system.

In summary, we have reported a compact and hardware-efficient implementation of a quantum repeater node in a single long-lived MAQM.
The long lifetime combined with the multimode individual addressing capacity of the MAQM paves ways for important applications in realization of quantum repeaters, long-distance quantum communication, and large-scale quantum networks.


\begin{acknowledgments}
We thank W. Chang, Y. Yu and X.-M. Hu for stimulating discussions and E. Salim for the help on vacuum.
This work is supported by the National key Research and Development Program of China (2020YFA0309500,2016YFA0301902), the Frontier Science Center for Quantum Information of the Ministry of Education of China, and Tsinghua University Initiative Scientific Research Program. Y.-K.W. acknowledges in addition support from Shuimu Tsinghua Scholar Program and International Postdoctoral Exchange Fellowship Program.
\end{acknowledgments}

\appendix

\section{Details of experimental setup}

The $^{87}$Rb atomic ensemble is prepared in a double-MOT setup. After a $79\,$ms loading stage, the ensemble is compressed for $10\,$ms, and is further cooled by polarization gradient cooling (PGC) for $7\,$ms, reaching a final temperature of $20\,\mu$K (details described in Ref.~\cite{PhysRevLett.124.240504}). The ensemble is then optically pumped to the ground state $|g\rangle=|5^2S_{1/2},F=1, m_F=-1\rangle$: the cooling beams are still on for $70\,\mu$s to pump the atoms from the $F=2$ to the $F=1$ hyperfine states; and another optical pumping beam resonant to the transition $|5S_{1/2},F=1\rangle\rightarrow|5P_{1/2},F=1\rangle$ with $\sigma_{-}$ polarization is also used to transfer the atoms from $m_F=+1$ and $m_F=0$ to $m_F=-1$ of the $F=1$ manifold. This optical pumping beam is kept on for another $30\,\mu$s after the cooling beams. After the optical pumping stage, the temperature of the ensemble slightly increases to $30\,\mu$K. A bias magnetic field of about $50\,$mG is applied to define the quantization axis throughout the whole process.

Different from the previous $4f$ configuration composed of two sets of AODs, two Fourier lenses and the ensemble \cite{pu2017experimental}, a couple of lenses are inserted into each of the multiplexing optical circuits (see Fig.~5). There are two purposes for these additional lenses: they increase the path length of the optical circuit to conveniently couple the Raman pumping beams to the multiplexing optical circuits; meanwhile they shrink the sizes of all the laser beams and the single-photon modes on the ensemble to allow the division of more memory cells without changing the switch time of the AODs. The Raman beams are initially coupled to the write/idler circuits to roughly set the direction; then the EIT storage method is implemented to precisely adjust the overlap of the Raman beams and the write/read/signal/idler modes on the ensemble (see detailed methods in Ref.~\cite{pu2017experimental}). A flip mirror and a camera help to calibrate the position of the Raman beams.

In the main text, the retrieval efficiencies of $5\times 5$ cells are measured. To address each of the cells, the RF frequency of the crossed AODs is swept from  $96\,$MHz to $112\,$MHz in the $X$ direction, and  from $100\,$MHz to $116\,$MHz in the $Y$ direction, both in a step size of $4\,$MHz. The maximal switch time of all the AODs is lower than $1\,\mu$s. When addressing several cells in the same row (column) coherently, we simply apply the corresponding frequency components on the $X$ ($Y$) AOD simultaneously with the desired amplitudes and phases. More details can be found in our earlier works, e.g., Ref.~\cite{pu2017experimental}.

\begin{figure}[tb]
  \centering
  \includegraphics[width=8.6cm]{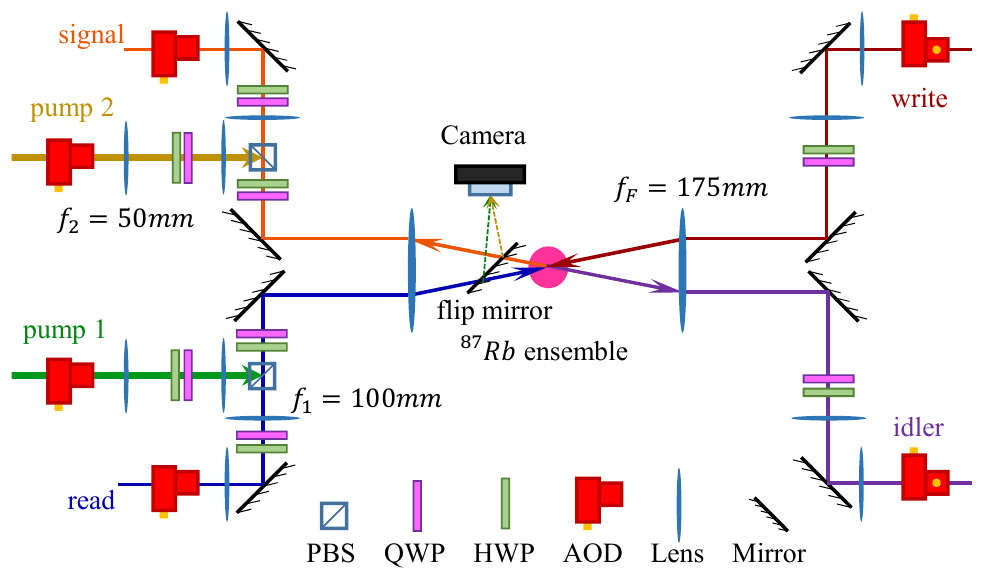}\\
  \caption{\textbf{} Detailed configuration of the setup.}
\end{figure}

\section{Pulse sequence}
After the ensemble is pumped to the ground state $|g\rangle$, a $500\,$ns clean pulse, identical to the read pulse, is applied to empty the storage state $|s\rangle$ of the target cell. A write pulse of 100 ns is applied to generate the signal photon and the correlated spin wave. If no signal photon is detected, a clean pulse is applied. Once a signal photon is detected, we apply the STIRAP pulses subsequently. The pulses are both in the Gaussian profile, with peak Rabi frequency $\Omega_1=\Omega_2=20\,$MHz. The pulse width should be long enough to satisfy the adiabatic condition, but also be short enough compared with the Larmor period to eliminate the momentum of the spin wave in time \cite{PhysRevA.84.043430}. The Gaussian pulse width $\sigma$ and the delay time between two pulses $\Delta t$ are optimized to obtain a highest retrieval efficiency, and the parameters used in the experiment are $\sigma=0.7\,\mu$s and $\Delta t=1.4\,\mu$s. The two Raman pumping beams are from the same laser so that their relative frequency and phase are stable. After the storage time $t$, the same STIRAP pulse with inverse order is applied, and a following $100\,$ns read pulse retrieves the spin wave into the idler photon. The probability of a signal photon detection is set to be $p_S\approx 0.4\%$ for all the experiments. Under this excitation probability, the anticorrelation, measured by splitting the idler mode into two equal parts to be collected by two SPDs, is $\alpha=0.145\pm0.014$.

\begin{figure}[tbh]
  \centering
  \includegraphics[width=8.6cm]{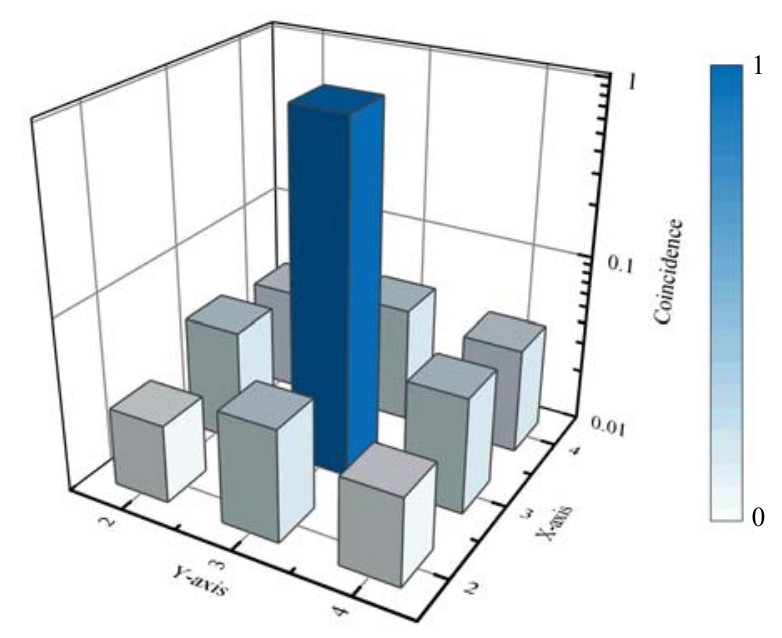}\\
  \caption{\textbf{} Crosstalk error at the storage time $t=10\,\mu$s.}
\end{figure}

\section{Imperfect STIRAP and crosstalk errors}

The distance between two cells is much larger than the size of write/read beams and the signal/idler modes, and the switch time ($< 1\,\mu$s) is smaller than the time interval between two read pulses ($5\,\mu$s), thus the crosstalk induced by a beam on an adjacent cell or the imperfect RF switching in AODs is negligible. Here, we measure the crosstalk originated from the STIRAP pulses. The measurement starts with generating a spin wave in the central cell and transferring it to the state $|\Psi\rangle_{gs^\prime}$. After that, the RF signals on the P1-AOD and P2-AOD are kept the same or switched to address an adjacent cell, and the STIRAP pulse is applied to the target cell after an additional $10\,\mu$s storage period. The retrieved idler photon counts are shown in Fig.~6. The maximal crosstalk is below $4\%$. This error might be due to the crosstalk of the STIRAP pulse, or the spin wave not fully transferred from $|\Psi\rangle_{gs}$ to $|\Psi\rangle_{gs^\prime}$. To verify this, we also measure the case when the storage time is $100\,\mu$s. The crosstalk errors are now well below $1\%$, which confirms that the main error source is the imperfect spin wave transfer by STIRAP pulses.

\section{Limitation on entanglement fidelity and retrieval efficiency}

The infidelity of the signal-signal entanglement comes from three aspects: the imperfect atom-photon entangled state generation, the decay during storage, and the error in Bell state measurement. The fidelity of the initial atom-photon entanglement is limited by the signal-to-noise ratio of the photon coincidence. When the excitation rate is too high, there will be large accidental coincidence between the signal and the idler photons which leads to a reduction in the measured cross-correlation $g_c$, defined as $g_c=p_c/(p_s p_i)$ with $p_s$, $p_i$ and $p_c$ being the probabilities to obtain a signal, an idler and a coincidence count, respectively. On the other hand, if the excitation rate is too low, the dark count of the detectors will become significant and will again lead to decrease in the measured $g_c$. Besides, lower excitation rates also result in longer measurement time to collect enough coincidence counts and the system may be subjected to stronger fluctuation. In this experiment, our excitation rate is limited by the measurement time where detection in each basis already takes 10 minutes to accumulate about 2000 counts. This gives $g_c\approx25$ and thus bounds the atom-photon entanglement fidelity $F\approx(g_c-0.5)/(g_c+1)=0.94$ \cite{de2006direct, PhysRevA.93.063819}. Then after $250\,\mu$s storage, the fidelity in the first link drops to $F\approx0.90$ when we generate the entanglement in the second link. Finally, the total infidelity of about $23\%$ after entanglement connection is a combination of the error in the two links together with that in the Bell state measurement.

The measured retrieval efficiency in Fig.~2b contains all the loss of the system. Specifically, we have a transmission efficiency of the AODs $\eta_{AOD}\approx0.95$, a fiber coupling efficiency $\eta_f\approx0.85$, an optical transmission efficiency including etalon filter $\eta_o\approx0.82$, and a detector efficiency $\eta_d\approx0.45$. From the measured retrieval efficiency of the central cell $R_m\approx 0.1$, we thus estimate the intrinsic retrieval efficiency between the write and the read processes as $R_i\approx0.35$. To improve the intrinsic retrieval efficiency, one promising solution is to use a cavity to enhance the atom-photon coupling \cite{Bao2012}.

%

\end{document}